# A Review on Recent Advances on Pool Boiling

by

**KAUSHIK MONDAL**

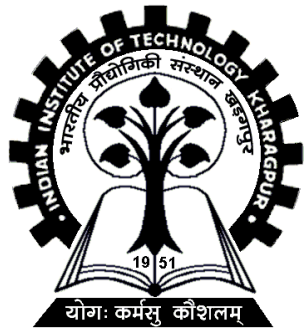

Mechanical Engineering Department.

Indian Institute of Technology Kharagpur

India

# Table of Contents

# Table of Figures

# 1. Introduction:

Pool boiling is one of the efficient way to remove heat from high power eletronics, heat exchanger and nuclear reactors. Now a days, pool boiling method is tried to exercise frequently due its ability to remove high heat flux compared with natural/forced convection, while maintaing at low wall superheat. But this pool boiling heat transfer capacity is also limited by an important paramenter, called critical heat flux (CHF). At the point of CHF, the heat transfer coefficienct (HTC) is drastically decreases due to the change of heat transfer regime from nucleate boiling to film boiling. This is why, the enhancement of CHF, while maintaining low wall superheat is of great interest to engineers and researchers.

Past studies on pool boiling heat transfer have shown that the enhancement of CHF depends on extended surface area, nucleation site density, wettability, capillary wicking. Keeping in mind of these factor, several attempts have been made to incraese CHF while decreasing the wall superheat at the onset of nucleate boiling. In many research work, it has been tried to enhance the CHF by adding metal foam [1], wire meshes [2], porous coating (with uniform or modulated structure) [3,4], roughnees [5] , and different structures [6] on top of the heated surface. Along with these surface modification, the effect of nano-particle [7,8] and surfactants [8] mixed solution on HTC in pool boiling heat transfer have been studied. Out of all these techniques of improvement of CHF and heat transfer coefficient, high porosity aluminium foam is chosen for enhancement of the same in this work. Increased surface area, nucleation sites and improved capillary action are turned to be the supported reasons for choosing such foam. Due to its large number of nucleation

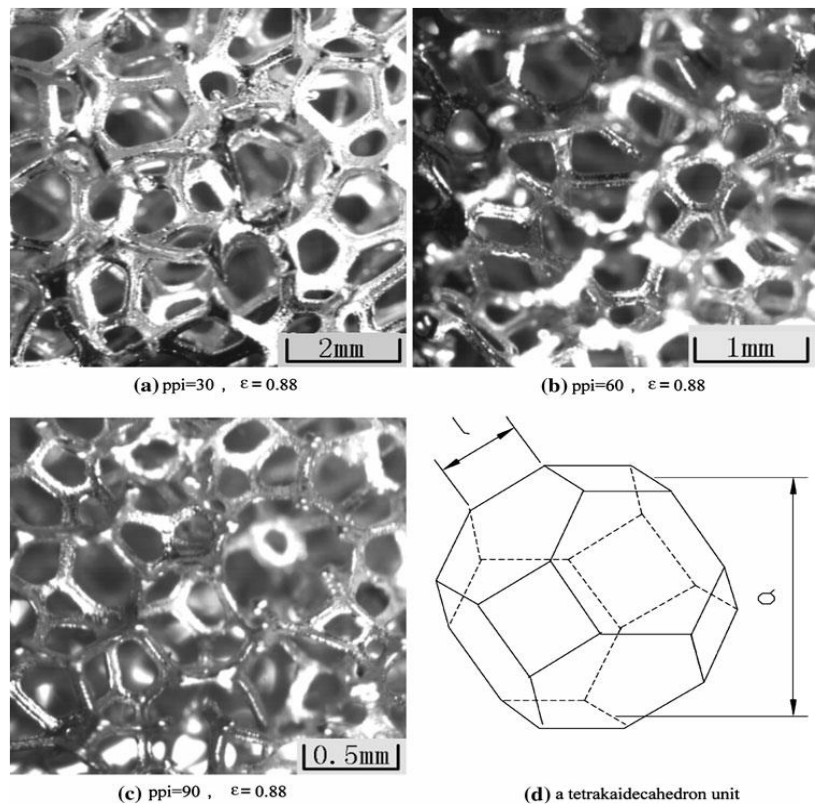

**Fig. 1** The photos of a metal foam (a, b, and c) and a unit foam cell representation (d) [12]

sites generate more numbers of bubble than plane metal plate. At the same time, smaller pore diameter tend to enhance capillary action, which is the main reason for replenishment of fresh liquid on heated area of the plate. Due to this large number of bubble generation, heat removal rate is seen to rise by a significant ammount compare with that of the same on a plane flat plate. Another important point, which is needed to be raised here is that the wall superheat at the onset of nucleate boiling (ONB) , is reduced while metal foam is used. It is shown in Fig. 2. that nucleate boiling has been observed at wall superheat of somewhere about

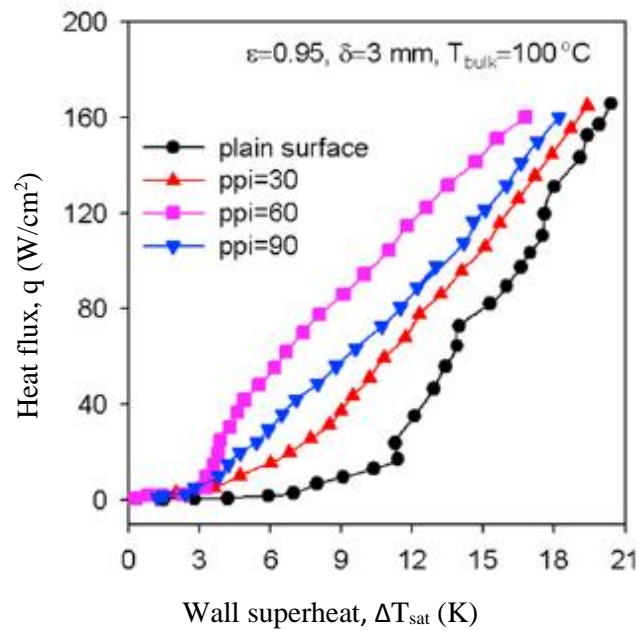

**Fig. 2.** Effect of metal foam of different ppi on wall superheat at ONB [18]

12 K on plane surface, whereas it is seen to be happened at the wall superheat of about 2-3 K on metal foam attached surface. There is another significant point to be noted here is the wall superheat is observed to be higher for plane surface in all the rage of heat flux.

## 2. Literature Review:

Li et al. [3] have done some experiments to show that modulated porous structures can drastically improve the critical heat flux (CHF) and heat transfer coefficienct of nucleate boiling heat transfer in de-ionized (DI) water. Delay of onset of hydrodynamic instability and enhancement of capillary pumping to replenish liquid in horizontal and vertical direction may cause in the improvement of heat transfer coefficient and CHF. In that work, a thick uniform porous structure, two modulated porous structures and two hybrid modulated porous structures have been used. A stable heat transfer coefficient of 2 times and CHF of 1.6 times are obtained respectively for uniform porous structures, compared with a plane surface. More active nucleation and extended surface area of uniform porous structures are cited to be the reasons for its improved performance. Then the next explored point is that the use of four porous pillars on a uniform porous base tends to increase the said parameters by 3 and 2 times to that of a plane and uniform porous structures respectively. Porous base of the modulated porous structures and prticle size used to make the porous structures are seemed to play very important role in the improvement of heat transfer coefficient and CHF.

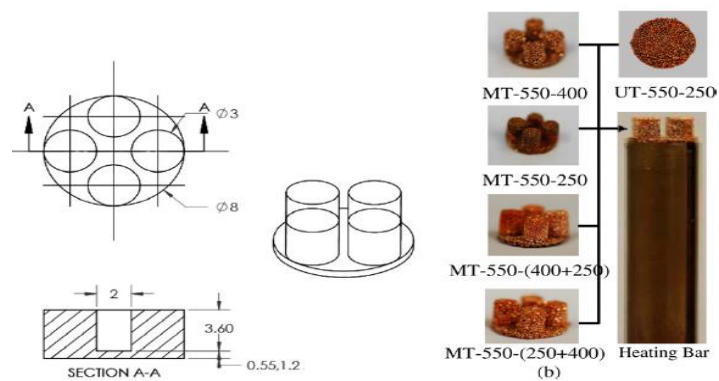

**Fig. 3.** Schematics of porous and photograps of sintered sample pictures. (a) Geometric dimensions of base and pillars of modulated porous structures; (b) the photographs of uniform thickness and modulated porous strcutures mounted on the heat bar.

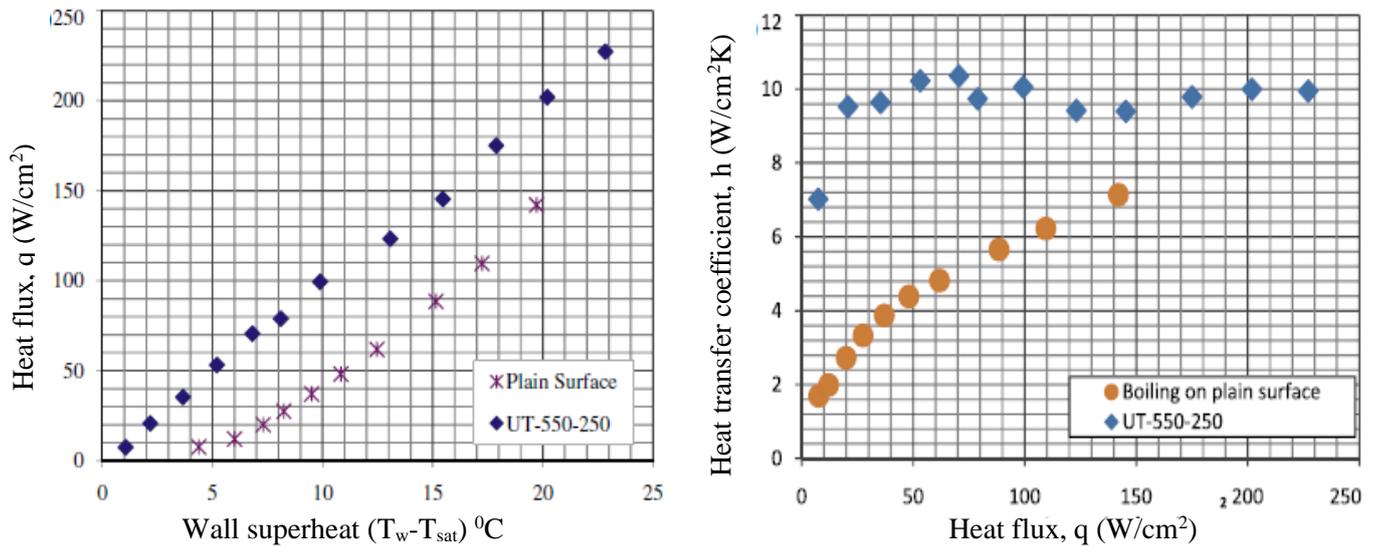

**Fig. 4**. Boiling curves of pool boiling on a plain surface and UT-550-250.

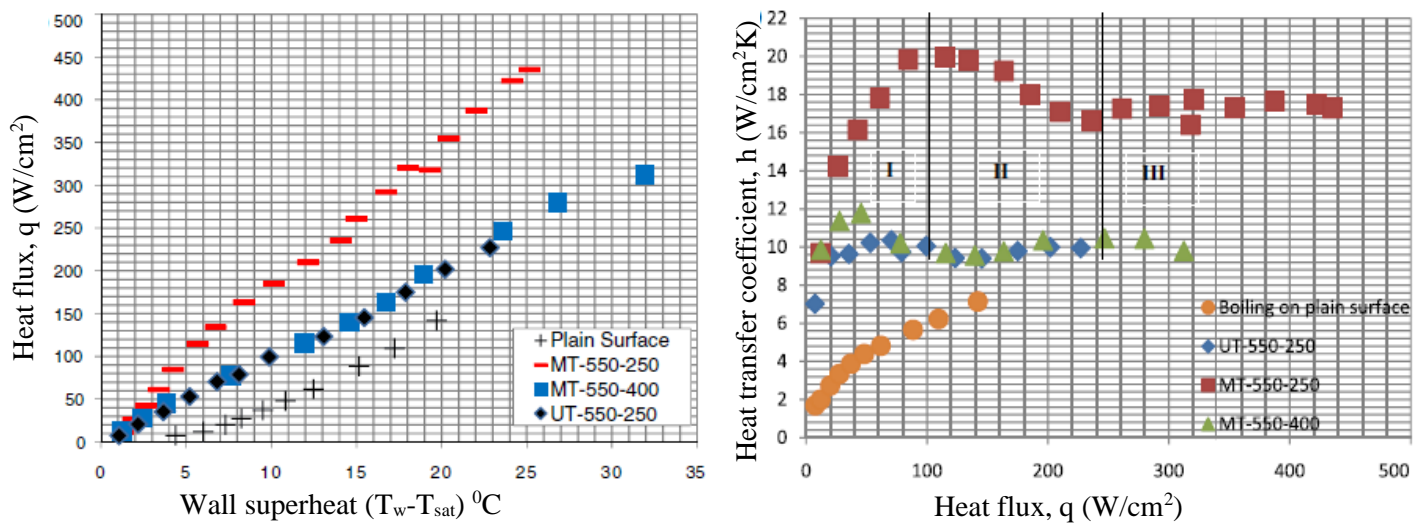

**Fig. 5**. Boiling curves of pool boiling on a uniform and modulated porous structures.

Capillary effect for liquid replenishment, surface area, bubble nucleation density and bubble departure frequency are observed to be closely associated with pore size. Delaying of onset of hydrodynamic instabillity and replenishment of vertical liquid have been significantly controlled by the porous pillars, whereas horizontal liquid replenishment hase been done by porous base of the modulated structure.

Launay et al. [9] have used micro- nano structures hybride surfaces in PF 5060 (flurionated fluid) and deionized water to investigate the performance of boiling heat trannsfer in the range of saturation temperatures from $35^0$C to $60^0$C. During experiments, various types of microstructures are used, e.g. smooth and rough silicon surfaces, with and without of carbon nanotubes (CNTs) as coat, etched silicon with CNTs made pin fin arrays and 3D microstructures. Silicon material and micromachining process are used to make all these microstructures. The experimental results are plotted in three graphs, e.g. maximum heat flux vs. saturation temperatures for PF 5060, heat flux vs. wall superheat at a constant saturation temperature ($T_{sat} = 45^0C$) for PF 5060 and maximum heat flux vs. saturation temperature for water boiling on different heat transfer enhancement structures. It is seen in the graph (a) that maximum amount of heat is dissipated from the test surface when the specified 3D structure is attached with it compared with any other surface modification mentioned earlier. Increament in exposed area for heat transfer by the 3D structure is conceptualized as the reason of higher heat dissipation capacity of the same. Plotted results of the experiment in the graph (b) revealed that the heat transfer from CNT coated micro and nano structured surface is lesser than the 3D micro structures and Si etched pin fin array. However it can be said that the use of micro and CNT based nano structured Si surface improve heat transfer, but at the same time its performance is seemed unsatisfactory for poor thermophysical properties of PF 5060. The last graph (c) of their work clearly conclude that the 3D enhanced microstructures outperform other micro and nano structure surfaces. It is also evident in the graph (c) that the flux of heat dissipation is much more in this case than it was for PF 5060 due to better thermophysical properties of the boiling fluid.

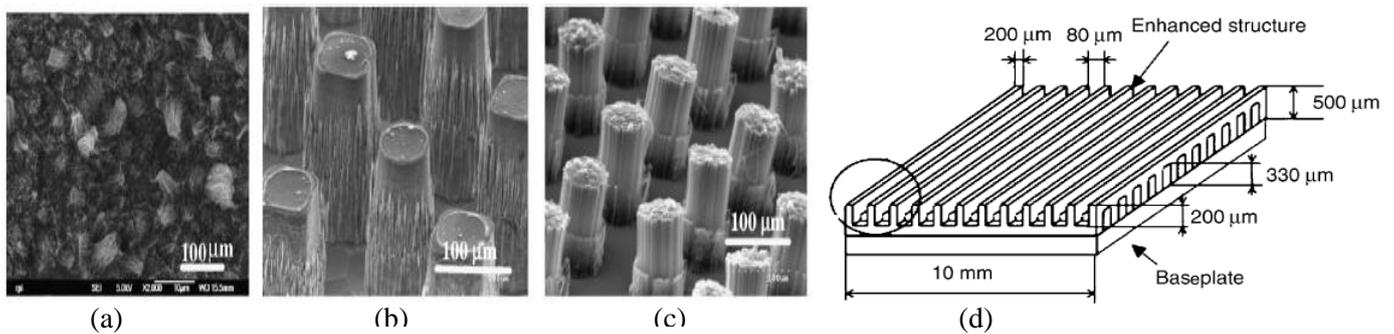

(a)  (b)  (c)  (d)

**Fig. 6**. (a) SEM view of a sample fully coated with carbon nanotubes. (b) SEM view of the silicon pin-fins structure. (c) SEM view of the carbon nanotube pin-fin array. (d) Schematic of 3D enhamcement microstructure.

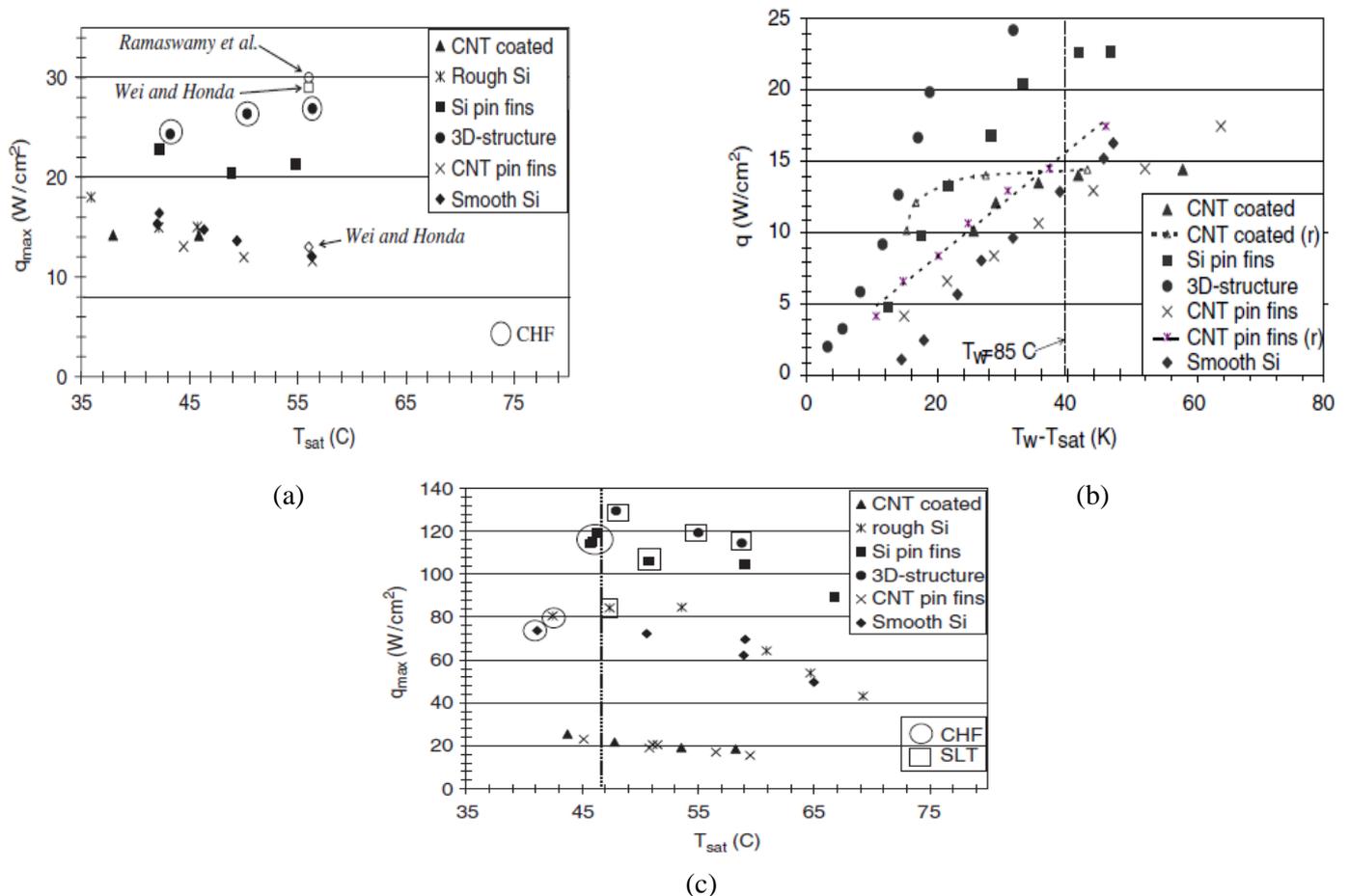

**Fig. 7.** (a) Maximum heat flux vs. the saturation temperature for PF5060 boiling on different heat transfer enhancement structures. (b) Boiling curves for PF5060 at $T_{sat} = 45^0$C. (c) Maximum heat flux vs. saturation temperature for water boiling on different heat transfer enhancement structures.

Mori et al. [10] used a honycomb porous structure (made with ceramic) on a heated surface to see it's effect on heat transfer in water under saturated condition. CHF is seen to be increased by 2.5 times for the use of this porous structure in comparison with plane heated surface. Creation of seperate flow path for liquid and vapour is idealized to be the reduction of possibile of instability, which resulted in delaying the film boiling phenomenon. Optimal size of pores opening is also found out for a specified thickness of the plate at which CHF occures.

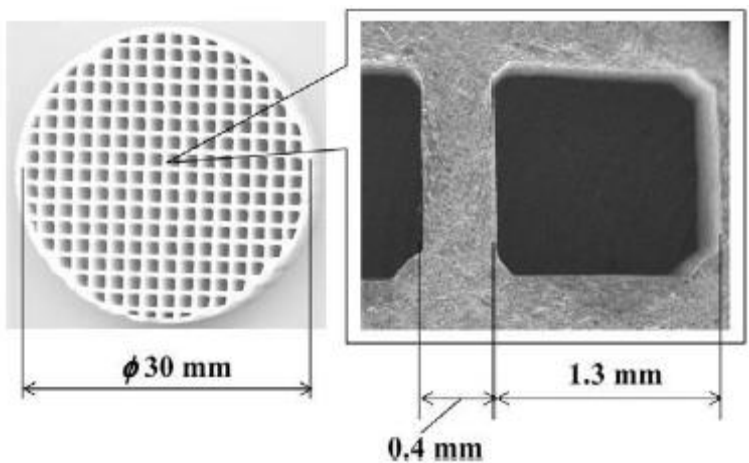

**Fig. 8.** Shape of honeycomb porous plate

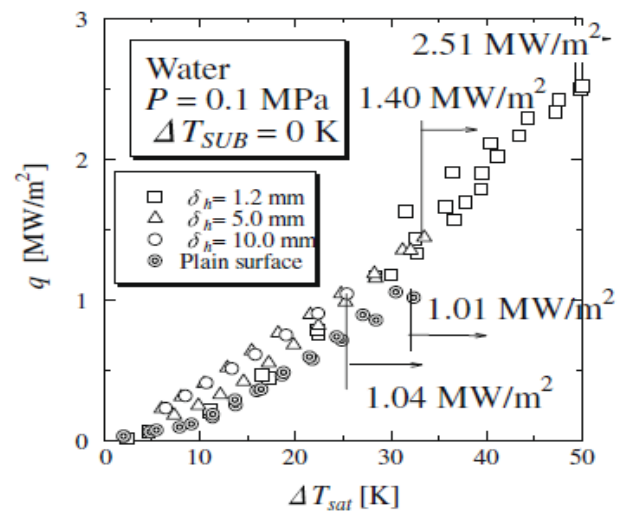

**Fig. 9.** Boiling curves of honeycomb porous plates of different height

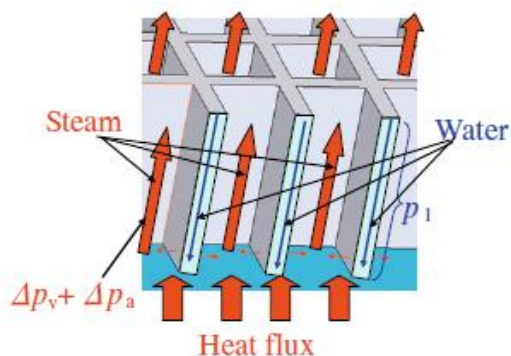

**Fig. 10.** Schematic diagram of steam and water flows in a honeycomb porous plate

Jaikumar and Kandilkar [11] have exercised a different approach to see the effect on CHF and heat transfer coefficient by creating sperated liquid feeder channels (FC) and nucleating regions (NR) on a heated plate.

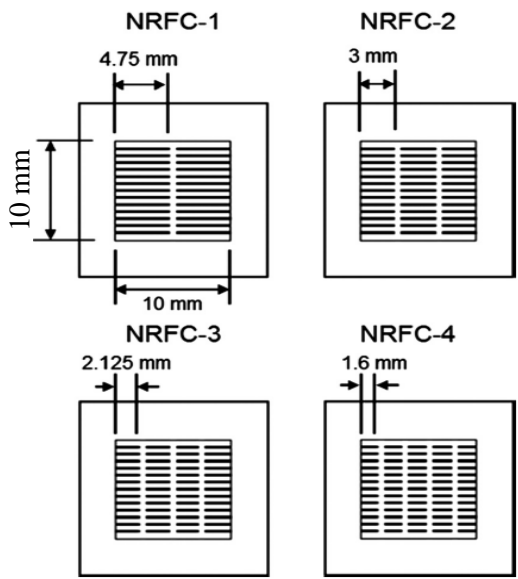

**Fig. 11.** Test chips used in this study

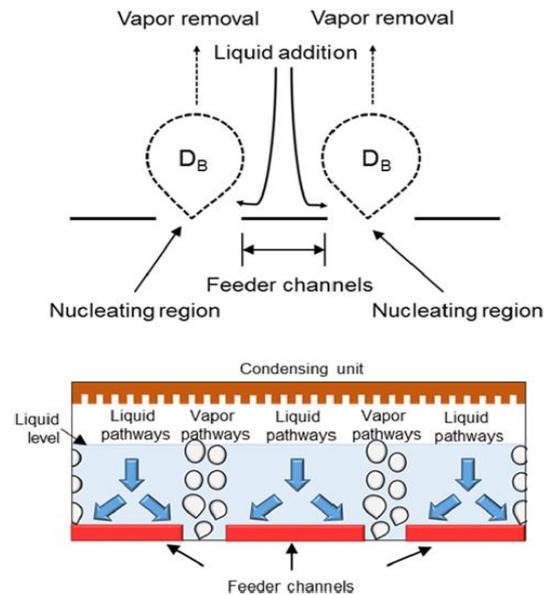

**Fig. 12.** Top-Schematic showing the liquid supply and vapour removal pathways for an NRFC configuration

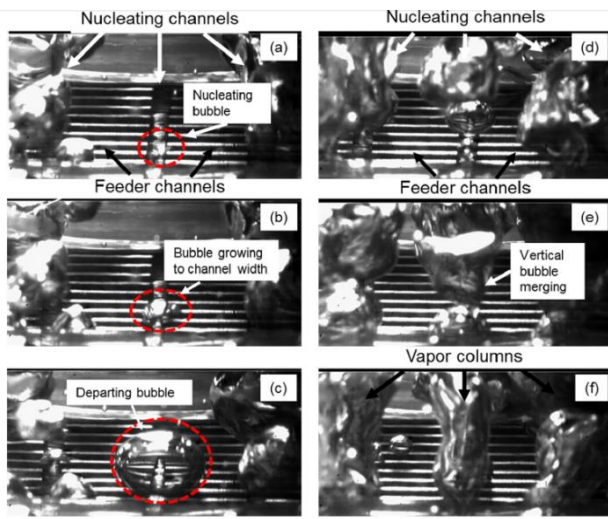

**Fig. 13.** Bubble sequence obtained with NRFC-3 surface

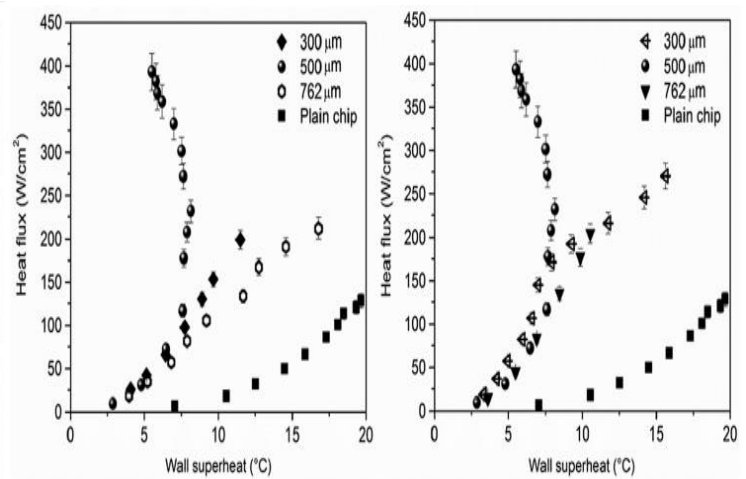

**Fig. 14.** (a) Effect of NR channel width. (b) Effect of FC channel width

The principle of vappour bubble induced convection is used to push the surrounding liquid towards the nucleating region through FC. The flow of freash liquid through the passage of FC is almost ramain unimpeded by the nucleate bubbles, created on the same path. The suppression of incoming liquid on nucleate bubbles, which are created on FC path is asumed to be the main reason for resistance free movement of liquid. CHF and heat transfer coefficient are significantly improved due to this said modification on heated surface. Width of FC is turned out to be a major factor through prametric studies of the modified surface in maximizing the said parameters. Heat transfer coefficient and CHF are seen to be maximized for a particular value of FC width, when it equals to the buble departure diameter. Furthermore, with the increase or decrease in FC width from that particular value, heat transfer coefficient and CHF are getting decrease.

Yang et al. [12] have experimentally investigated the effect of inclination angle of a heated surface attached with copper foam. It was seen that the use of copper foam reduces surface superheat significanly at the onset of nucleate boiling (ONB) and widen range of surface superheat by increasing the CHF, compared with plane smooth surface. Boiling curves are seemed to be crossed between low and high inclination angle, and it was also observed that thermal performance decreases with increases in inclination angle at small or moderate surface superheat or heat flux, but thermal performance increases with the increasing value of inclination angle at higher surface superheat or heat flux. The reason of this interesting results is explained as follows. At small or moderate heat flux range, the generated bubbles are more freely escape from horizontally positioned heated surface than vertically positioned surface, because of smaller bubble detachement resistance. Although at the larger heat flux region, vertically positioned heated plate produces significant buoyancy force to suck liquid from the bottom of the plate, which tends to improve its performance at high heat flux region. The effect of foam cover thickness (3.0, 4.0 and 5.0 mm) on heat transfer coefficient has been explored in this study, where the outcome of the results in Fig. 16. show that upto the surface superheat of 20-25 K, maximum thicken copper foam outperform other due to more extended heat trasnfer area and nucleation site. But with the increasing value of surface superheat from the earlier specified surface superheat, heat transfer coefficients' of 3.0 and 4.0 mm thcikness metal foam outperform 5.0 mm thickness foam due to increasing vapour release resistance. In this context it is also important to mention that boiling incipence phenomenon is observed only when 3.0 mm thicken metal foam is used. This work has also explored that the effect of bulk liquid temperature on pool boiling heat transfer on metal foam covers. It is seen that pool boiling heat transfer is very insensitive to inclination angle at low bulk liquid temperature of $38^0$C. At this low bulk liquid temperature generated bubles are easily condensed in the foam, which make heat transfer insensititive to the inclination angle. Another part of the results, show that heat transfer performance is better at higher bulk liquid and low wall superheat condition due to easy bubble generation in foam, causing better heat transfer. But with the reduction of bulk liquid temperature and increase in wall superheat cause better heat transafer than previously stated condition. Reduction in vapour escaping resistance due to condensation of the genedrated bubble inside the foam is conceptualised to be the main reason. At the last of this work, one correlation is made among Nusselt number (Nu), Stanton

$$Nu=3.915\times10^{-3}St^{-0.214}Bo^{0.037}Re^{0.974}F^{0.038}ANG^{-0.067}\Theta^{-0.715}$$

number (St), Reynold number (Re), Bond number (Bo), pool liquid subcooiling effect, inclination angle effect, and more or less 20 percentage of deviation is found between calculated values from this correlation with the experimental results.

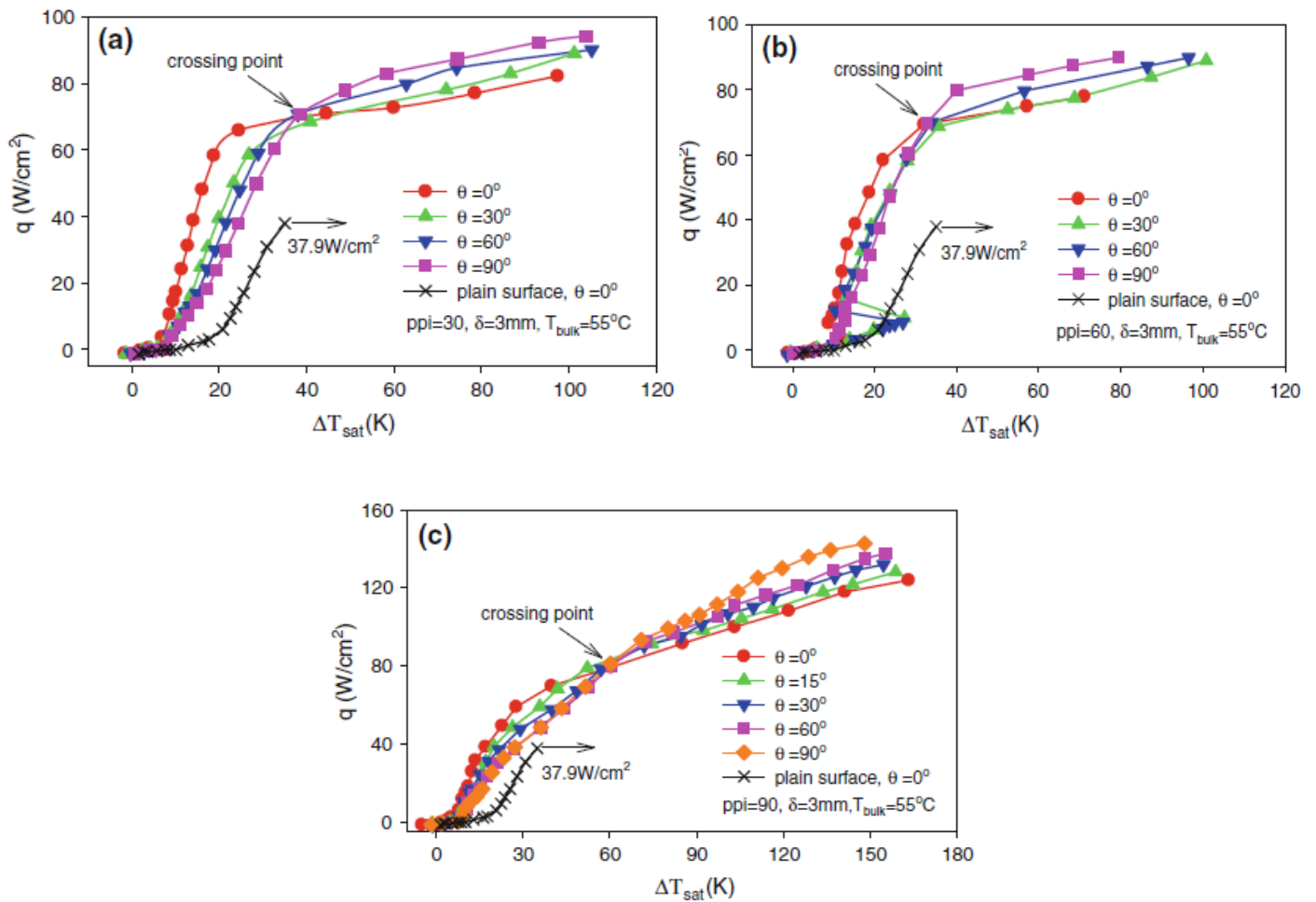

**Fig. 15.** Effect of inclination angle on boiling curves at pool liquid temperature of $55^0$C

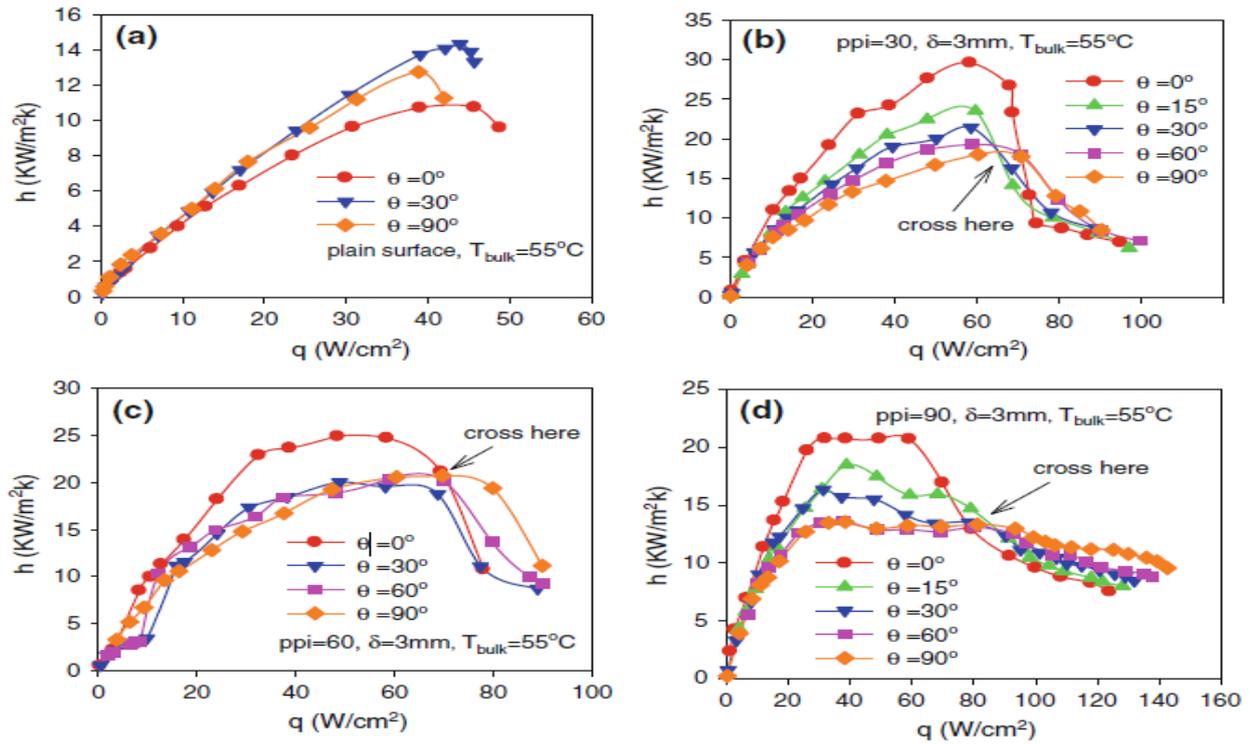

**Fig. 16.** Effect of inclination angle on heat transfer coefficient of plain surface and copper foam covers.

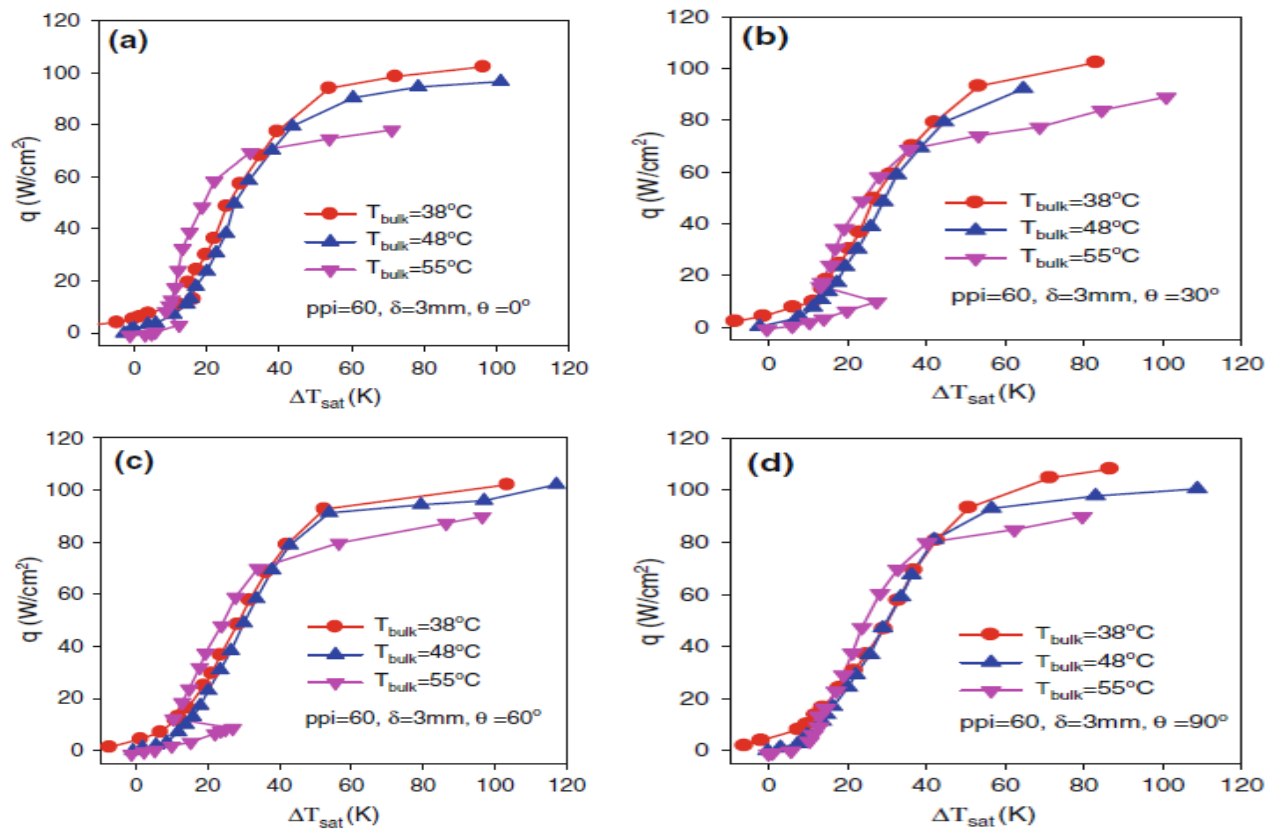

**Fig. 17.** Effect of pool liquid temperature on boiling curves at various inclination angles

Xu et al. [13] have done some research work on saturated pool boiling heat transfer in deionized water added with surfactants on a horizontal copper foam surface (under a wide pore density range 20-130 PPI with porosity of 0.95) with V-shaped grooves. In this study, sodium dodecyl sulphate (SDS) is used as

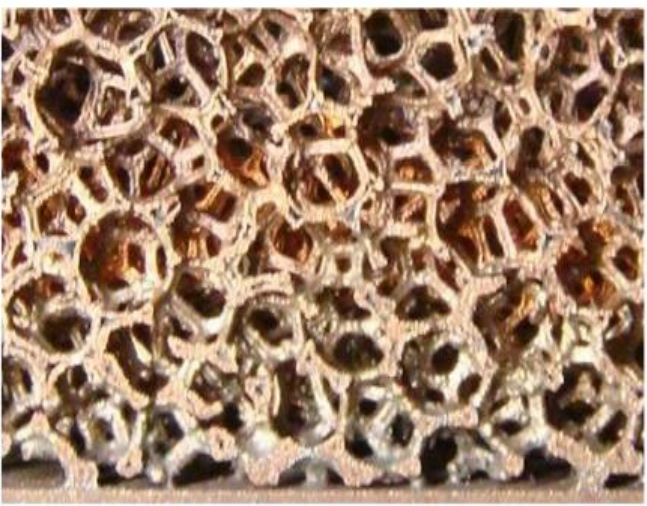

(b)

ith copper substrate. (b) Sketch of V-shape grooves.

surfactant. The effect of grooves on pool boiling heat transfer is found to be strongly dependent on pore density. At first, the effect of groove width with small pore density metal foam is investigated, where it is seen that CHF is considerably delayed with the use of grooved metal foam compare with plane metallic surface. Increament in number of nucleation sites, and suction of fresh water due to capillary action are considered to be main rasones of improve heat transfer performance of grooved structure metal foam. On the otherside, higher pore diameter of the low pore density foams reduces capillary force, which finally resulted in poor heat transfer performance with icrease in groove width. The effect of groove number on heat transfer performance in 100 PPI foam is tested, and observed to a crucial parameter for efficient parameter. It is observed that the foam of 100 PPI surface is almost covered with vapour bubbles with zero, one and two grooves, which resulted resistance to the fresh incoiming liquid from getting intuch with the heated surface. But with 3 number of grooves, almost all bubbles are escape out form the surface and fresh liquid is easily sucked towards the heated surface by the foam, and creates a steady circle .It is seen from the experimental results that the increasing number of grooves enhance heat transfer by reducing counter flow resistance between upward moving vapour and incoming liquid. Moreover, it is also observed that enhancement of CHF is proportional to the numbers of grooves for 100 PPI foam. In this manner, metal foam of 130 PPI pore density was also tested, and seen that the effect of grooves on its performance is not as signifiacnt as it is in for 100 PPI denstity foam. In the later part of this work, SDS concentration effect has been studied. It is seen that the heat transfer coefficient decreases with increasing SDS concentration. This is happen because mean bubble size becomes smaller, and their numbers increase sharply with increasing value of SDS concentartion, which may increase counterflow resistance signifiacntly. This phenomenon becomes very prominent in a higher heat flux region. During experiment, the effect foam cover thickness has also been captured. It is seen that foam cover thickness increases surface area for heat transfer but at the same time it enhances resistance for the vapour to leave from the foam. So one optimaization is done on groove free surface at a fixed pore density, which shows that lower pore density (higher pore diamter) gives higher value of optimum foam cover thickness and higher pore density (lower pore diamter) gives lower value of optimum foam cover thickness. Particularly, at high foam cover thickness made with copper gives low heat transfer coefficient compare with Nickel made foam due to bubble congestion inside the copper foam material.

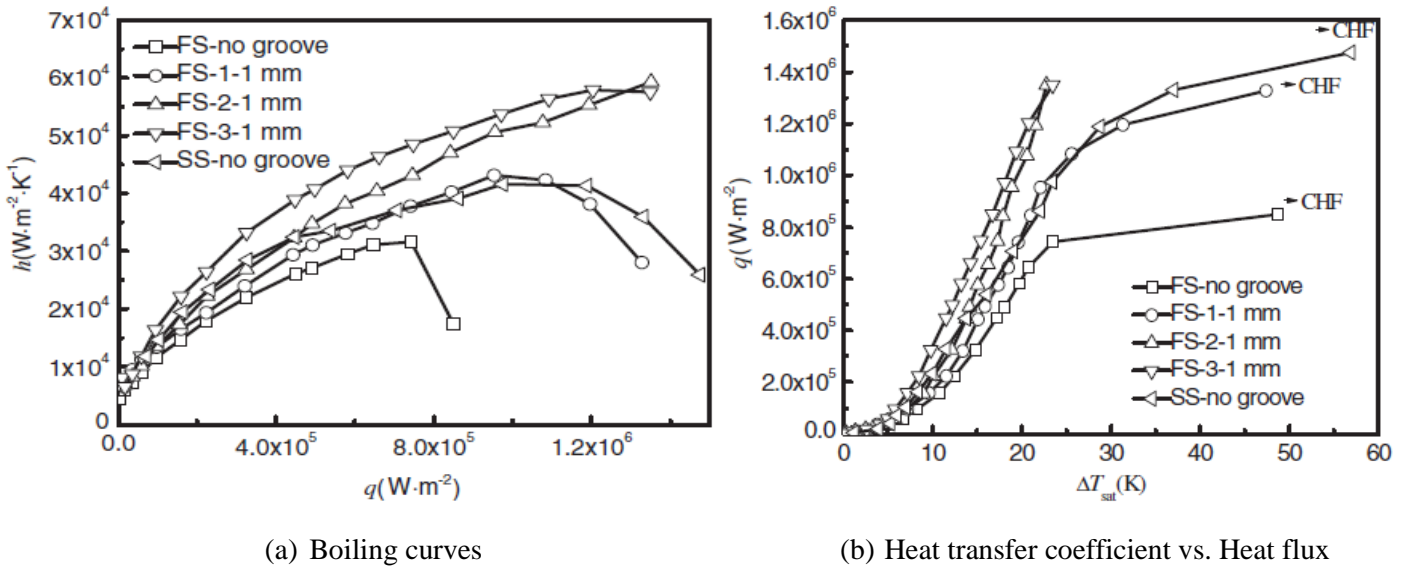

(a) Boiling curves  (b) Heat transfer coefficient vs. Heat flux

**Fig. 19.** Effect of groove width on boiling curves and heat transfer coefficients under the saturation pool liquid Conditions (100 PPI, $\varepsilon=0.95$, $\delta=3$ mm, W= 1 mm).

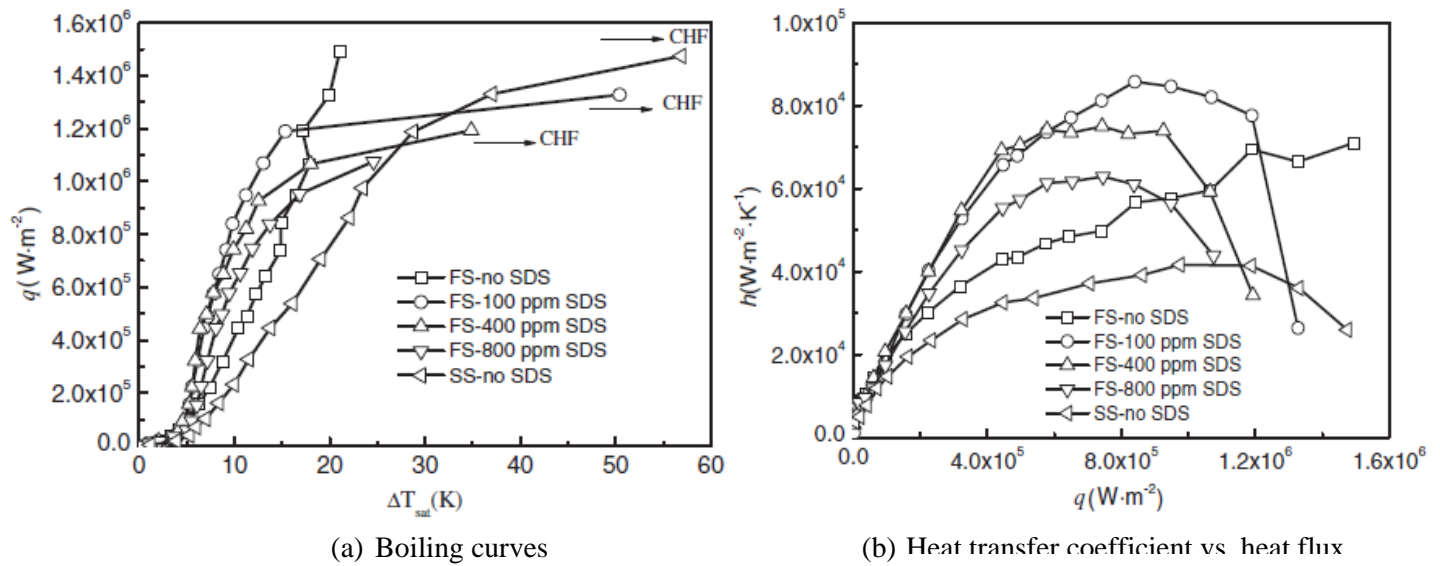

(a) Boiling curves  (b) Heat transfer coefficient vs. heat flux

**Fig. 20.** Effect of SDS on boiling curve and heat transfer coefficients under the saturation pool liquid conditions (20 PPI, $\varepsilon=0.95$, $\delta=7$ mm, W= 3 mm, D=10 mm, n =2)

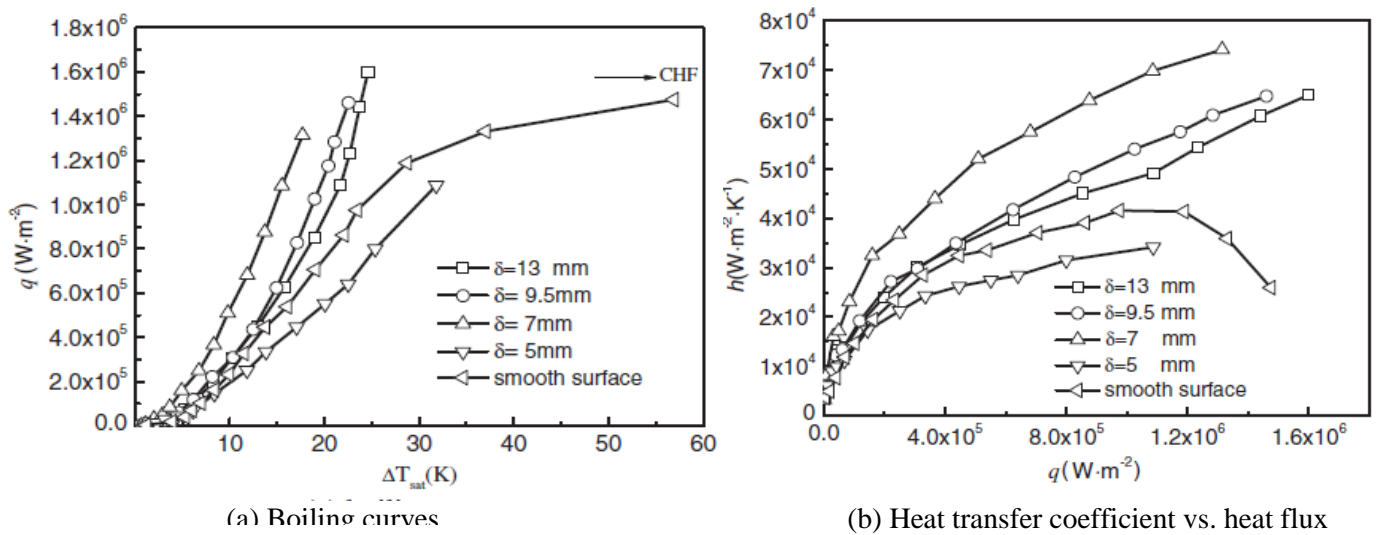

(a) Boiling curves  (b) Heat transfer coefficient vs. heat flux

**Fig. 21.** Effect of foam thickness on boiling curves and heat transfer coefficients under the saturation pool liquid conditions (20 PPI, $\varepsilon=0.95$).

Pranoto et al. [14] studied boiling heat transfer on pocofoam (61% porosity and kfoam 78% porosity). In this study two structured block and fin graphite foam evaporaters with different fin-to- block surface araea ratio

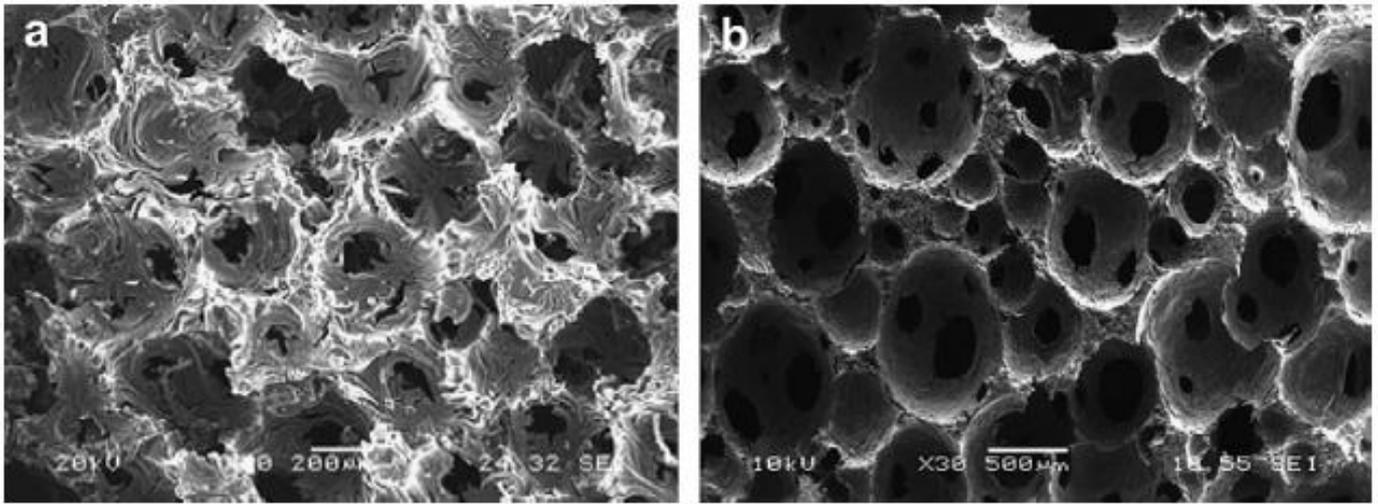

% porosity graphite foam and (b) "Kfoam" 78% porosity graphite foam.

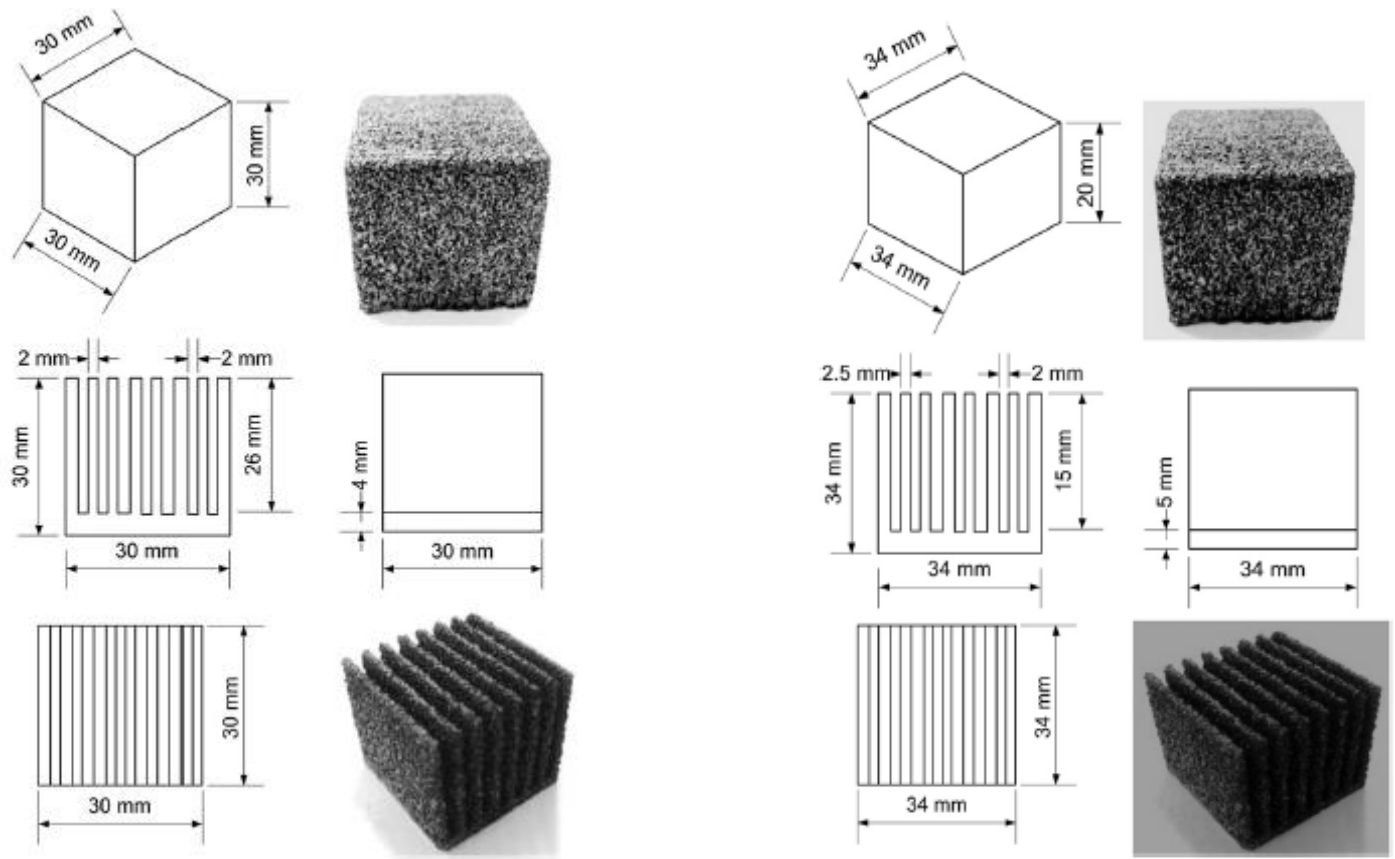

**Fig. 24.** Dimensions of graphite foam of block and fin structures for AR=2.73.

(AR) are tested in FC-72 and HFE-7000. The results show that the average wall temrperature difference between block and fin structures are found to be about $8^0C$ to $10^0C$ for AR= 3.70 and 2.73 respectively. The combined results of wall temperature and wall superheat on block and fin graphite foam revealed that the boiling performance is poorer for structured fin surface. For all the tested foams, coolants, and structures of different aspect ratios, the average boiling heat transfer coefficients of the block structure are about 1.2 to 1.6 times higher than those of the fin structure for the tested heat flux level of up to 112 W/cm$^2$. The results also indicate that the evaporator structures type significantly affect the boiling performance. The boiling images captured by a high speed camera supports the evidence that the fin structure reduces the number of nucleation sites and decreases the boiling performance compare with evaporator block. The study also identified that FC-72 causes higher wall temperature due to its higher saturation temperature, which resulted in improve boiling heat transfer performance compared with HFE-7000. The Bo, Ca, and Gr are calculated for HFE-7000 and compared with the values of the said non dimensional number for FC-72. Their comparative results indicate that the large ratio of buoyancy to surface tension force in FC-72 promoted bubble growth and departure, resulting in higher bubble departure frequency.

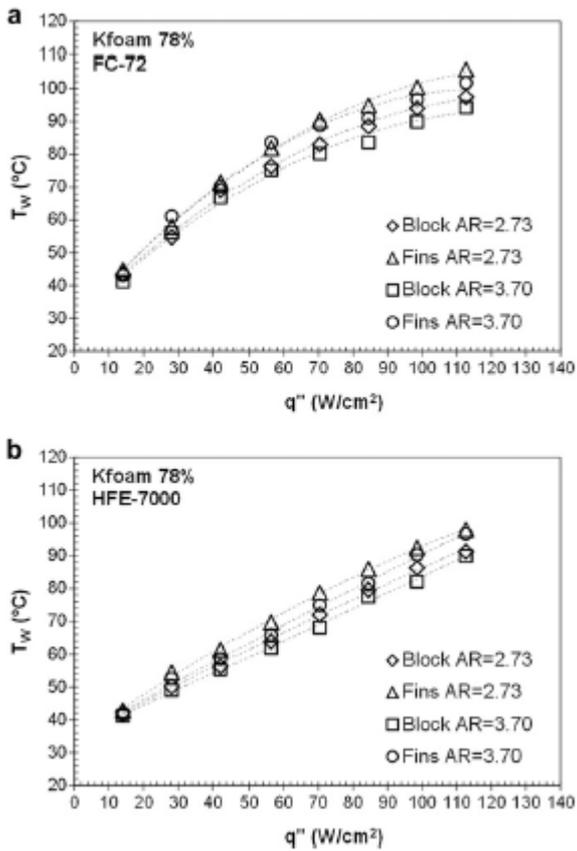
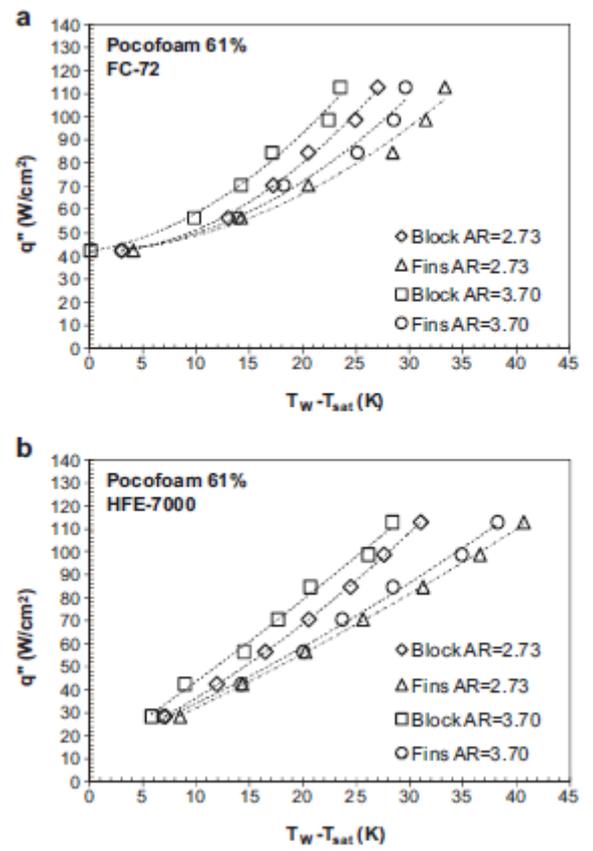

**Fig. 26.** Boiling curves of "Pocofoam" 61% porosity graphite foam in FC-72 and HFE-7000 for different block and fin structures.

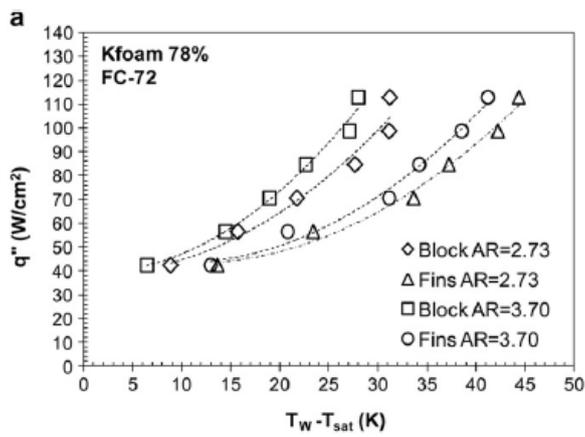
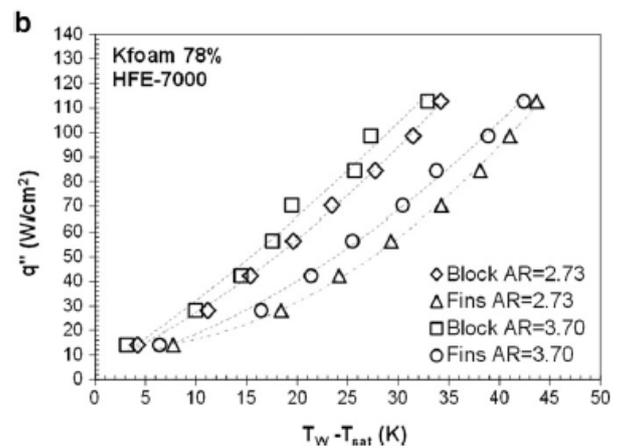

**Fig. 27.** Boiling curves of "Kfoam" 78% porosity graphite foam in FC-72 and HFE-7000 for different block and fin structures.

Xu and Zhao [15] have investigated pool boiling heat transfer performance using gradient pore metal foam, in nanoparticle and surfactants mixed de-ionized water at atmospheric pressure. Copper and nickel were used as foam material in the experiment. The pore densities of the foam layers are chosen as 5 PPI and 100PPI with constant porosity of 0.98. The gradient nickel foam which has been chosen for experiment consists of 20 PPI and 100 PPI down foam layer with the thickness of 2 mm and 5 PPI up foam layer with

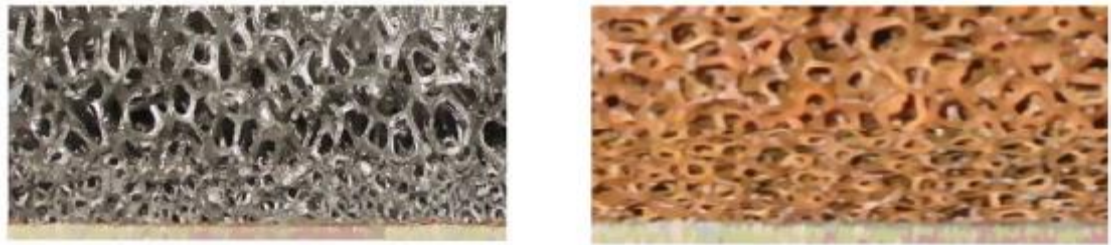

(a)            (b)

**Fig. 28.** Images of (a) a gradient nickel, (b) a gradient copper foam

the thickness varies from 4 to 8 mm. Experimental results have proved that surface area increments by sintering upper foam layer of 5 PPI pore density with different thickness on the bottom foam layer of 20 PPI pore density has very less effect on pool boiling heat transfer performance. Although another aspect of this study revealed that when 6 mm thicken bottom porous layer of (20 PPI & 5 PPI) gradient metal foam is

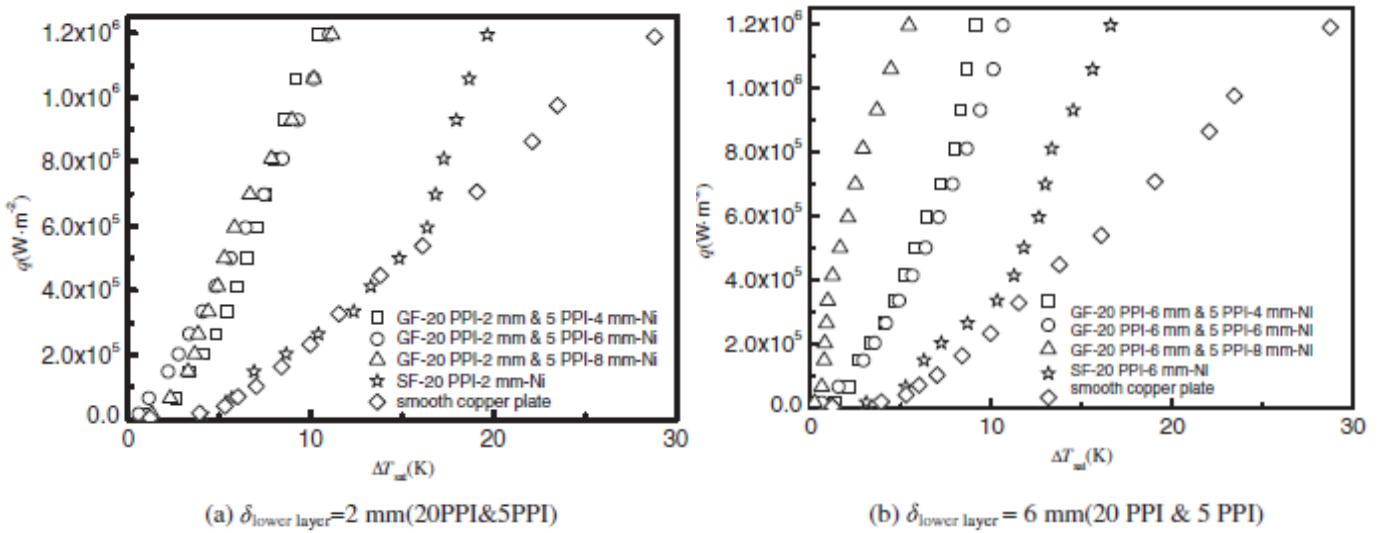

(a) $\delta_{lower\ layer}$=2 mm(20PPI&5PPI)      (b) $\delta_{lower\ layer}$= 6 mm(20 PPI & 5 PPI)

**Fig. 29.** Upper foam layer thickness effect

sintered with various thicknesses 5 PPI metal foam, initially wall superheat increases with increase in heat flux, later the surface superheat decreases with increasing value of surface heat flux. It was observed that upto a certain heat flux limit, pool boiling heat transfer increases with increasing thickness of bottom metal foam layer. As the imposed heat flux exceeds that particular limit, heat transfer performance starts falling. This kind of phenomenon is observed due to increasing value of vapour escaping resistance from the metal foam at high heat flux region, causes obstruction to the incoming fresh liquid from coming in contact with the heated surface. During the experiment, when alumina is used as nanoparticle, it worsens the boiling heat transfer performance because it reduces effective thermal conductivity of the porous medium. With the addition of SDS surfactant, pool boiling heat transfer performance is seen decreasing with increasing amount of SDS concentration.

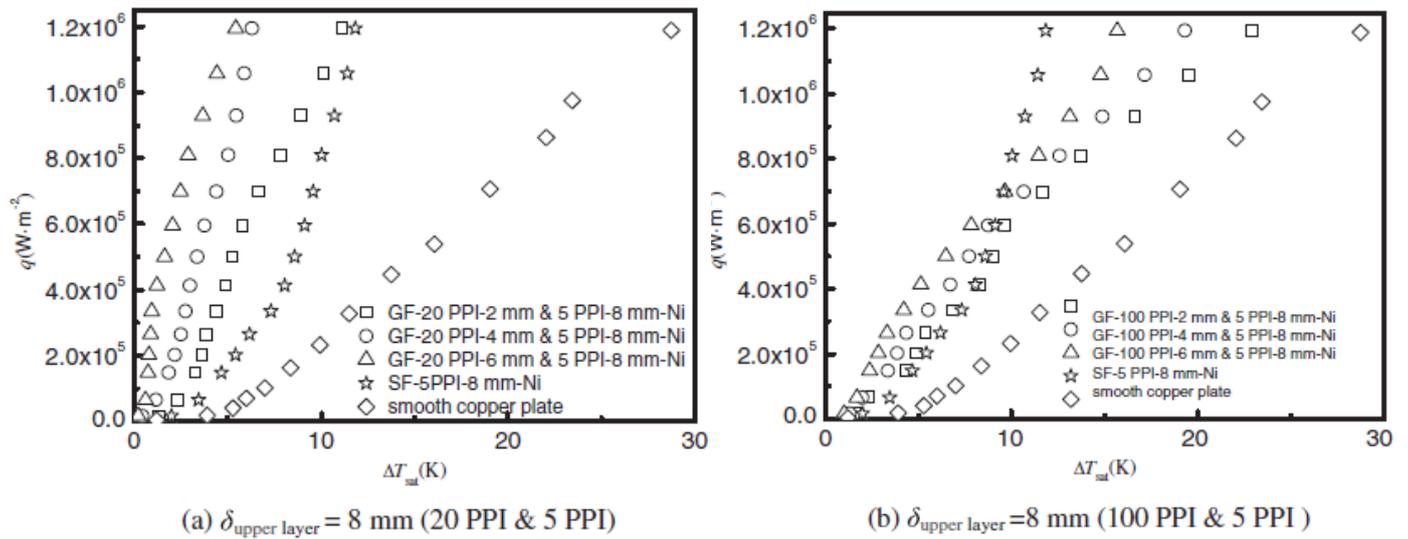

**Fig. 30.** Lower foam layer thickness effect.

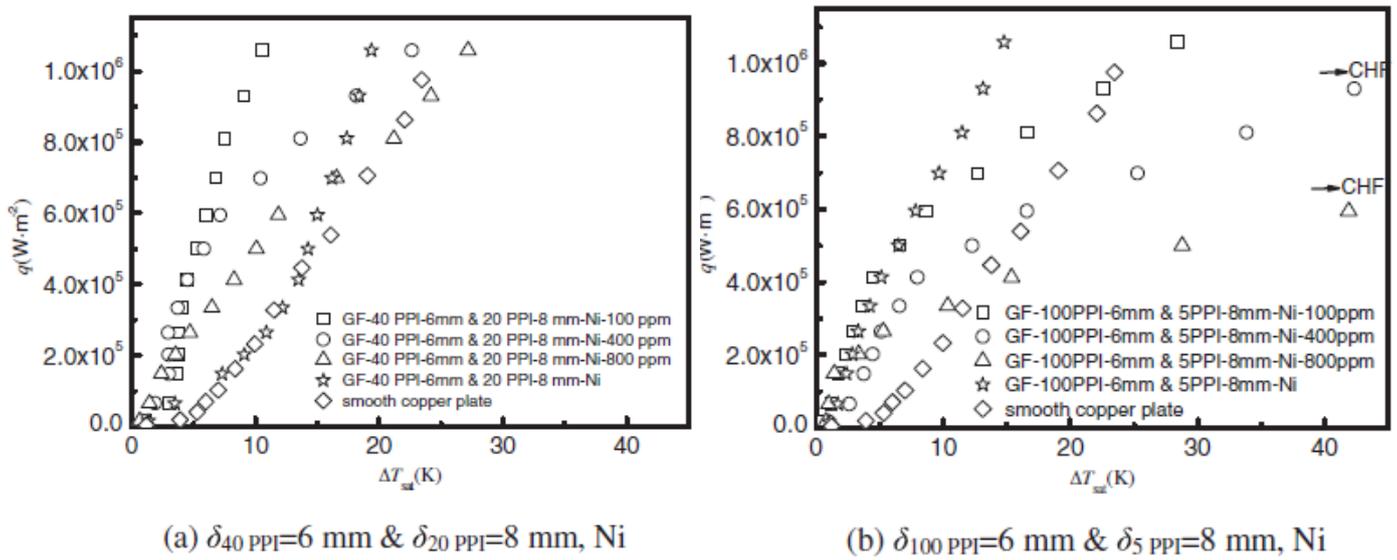

**Fig. 31.** Effect of SDS concentration

Deng et al. [16] experimented to see the effect on pool boiling heat transfer on porous coating and solid surface with re-entrant cavity. Porous coating is made with copper particles. At low and moderate heat flux region, porous coating with re-entrant cavity is performed well due to lower rate of bubble growth reduces the resistive force on incoming fresh liquid. However, at moderate to high value of heat flux, solid surface with re-entrant cavity starts performing well compare with porous coating due to increasing rate of bubble growth creates a clot of vapour embryo, which reduces the pool boiling performance of the porous coating. Although this reduction in pool boiling heat transfer coefficient at high heat flux region on porous coating is found to be much less as compared to its value, reported in other study. Fresh liquid supplied through the cavity and capillary action of the pores, are assumed to be main reasons for improved performance of the porous coating. On the other hand, unstable vapour column driven micro convection and larger thermal conductivity of the solid metal than the porous structure at slightly higher heat flux region are seemed to spread heat more efficiently from the heating surface to the surrounding region. Therefore solid surface shows higher heat transfer coefficient at slightly higher heat flux, compare with porous surface. Nevertheless, with further increase in heat flux a continuous and fairly stable vapour clot is seen to form above the surface. Unlike the porous coating, the absence of capillary action in solid surface (copper surface) fails to supply fresh liquid towards solid surface, which are imagined to reduce CHF. It is seen that the increase in subcooling level, the bubble nucleation is hindered due to the condensation, which tends to increase the wall superheat at ONB.

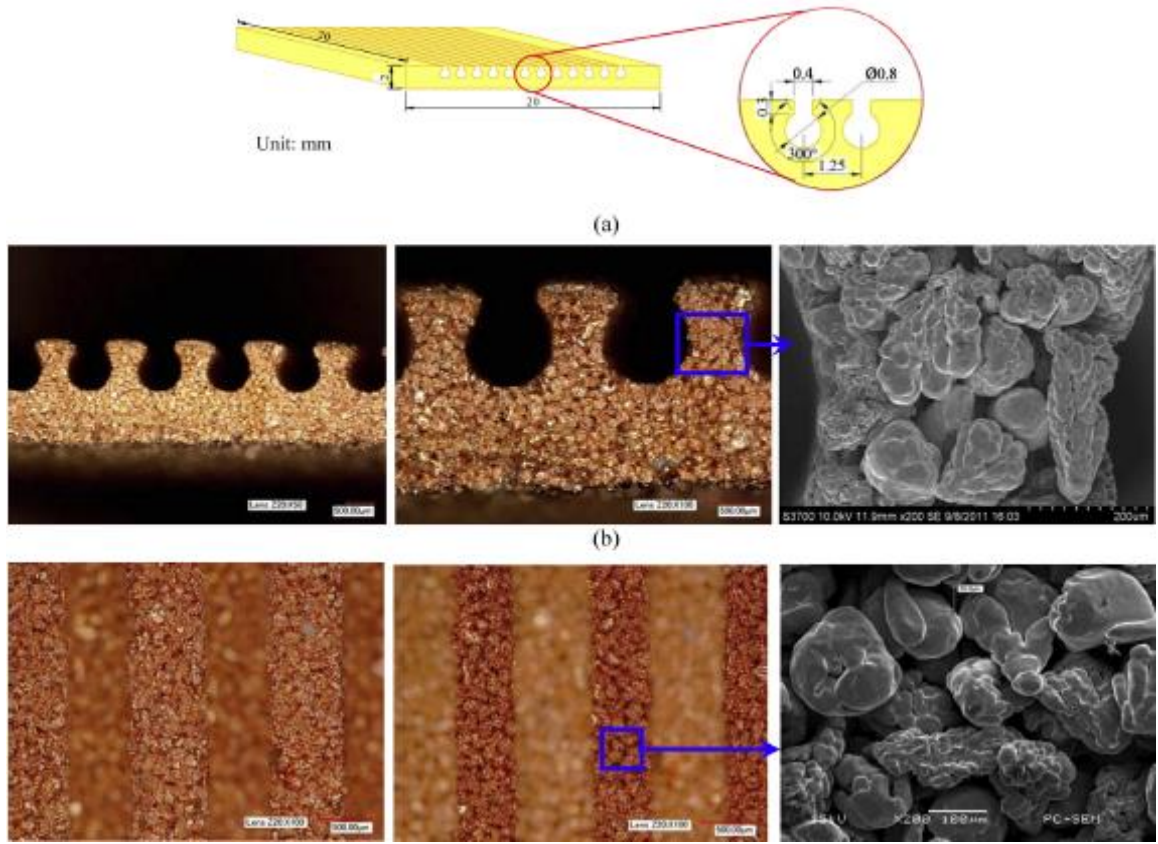

**Fig. 32.** Porous coating with reentrant channels; (a) geometric dimensions; (b) cross-section view; (c) top view: the top fins are focusing in the left picture, and the bottom wall surfaces are focusing in the middle one.

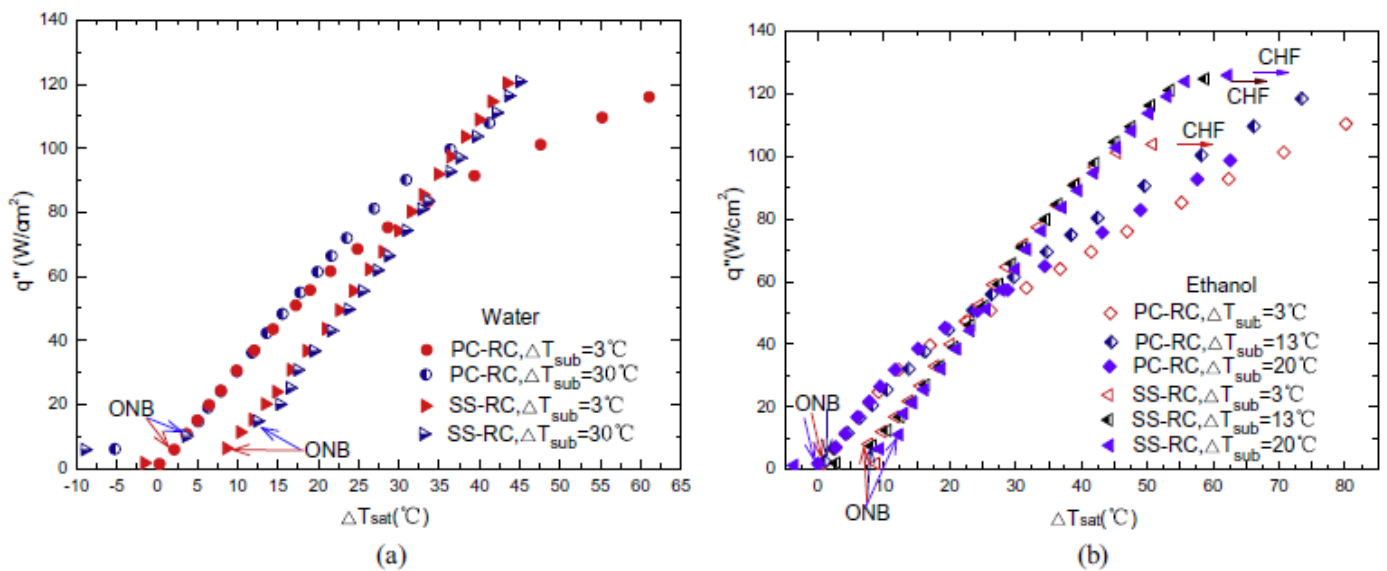

**Fig. 33.** Comparisons of boiling curves between porous coatings (PC-RC) and solid structures with reentrant channels (SS-RC): (a) water tests; (b) ethanol tests.

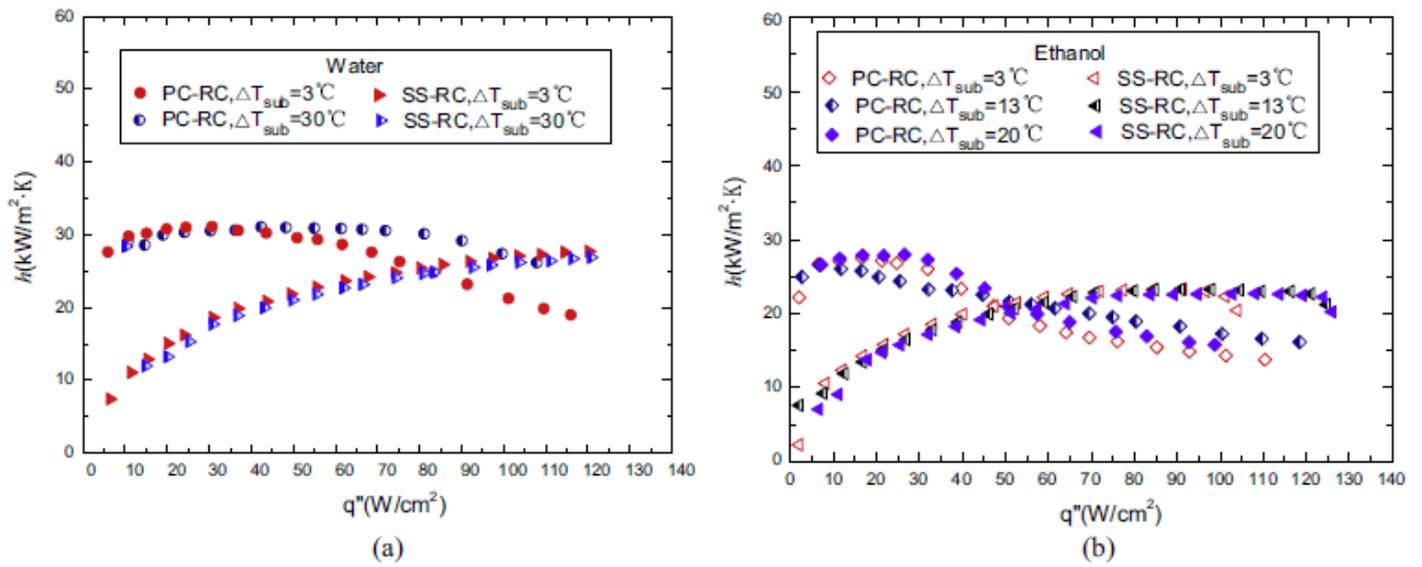

**Fig. 34.** Comparisons of heat transfer coefficients between porous coating (PC-RC) and solid structures with reentrant channels (SS-RC): (a) water tests; (b) ethanol tests.

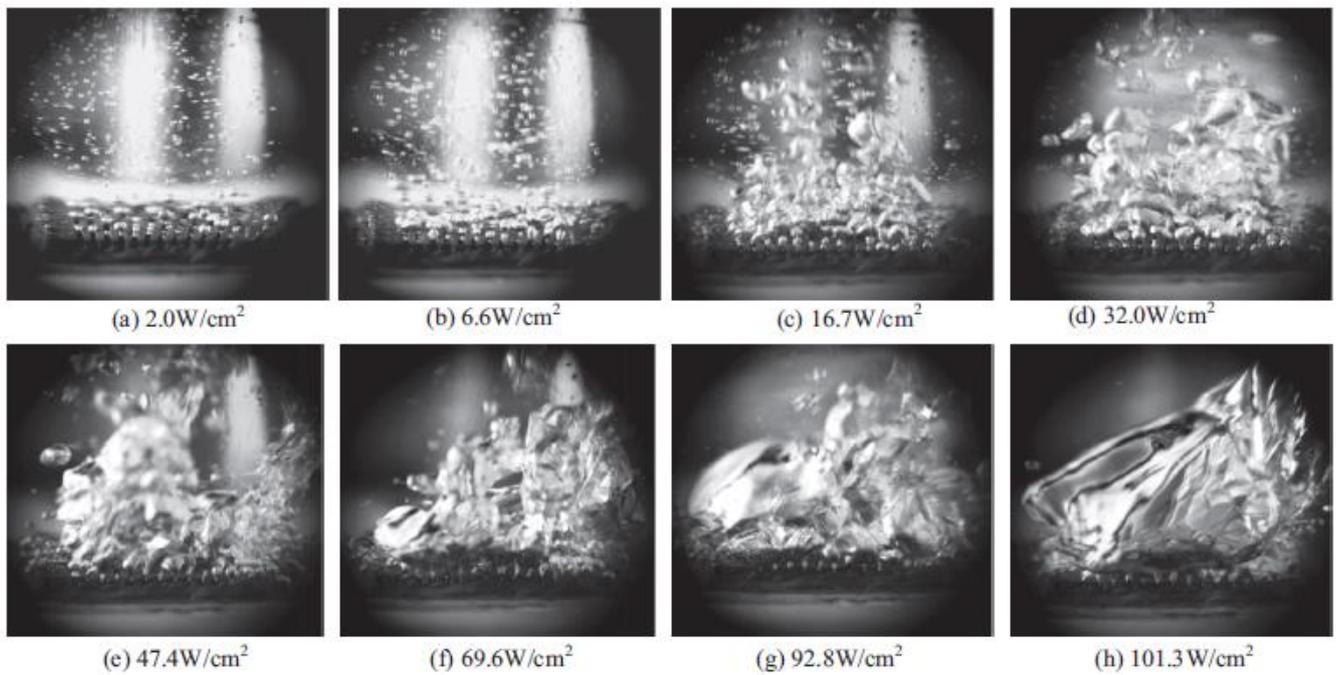

**Fig. 35.** Boiling visualizations of porous coating with reentrant channels ($\Delta T_{sub} = 3\ ^{0}C$, ethanol tests).

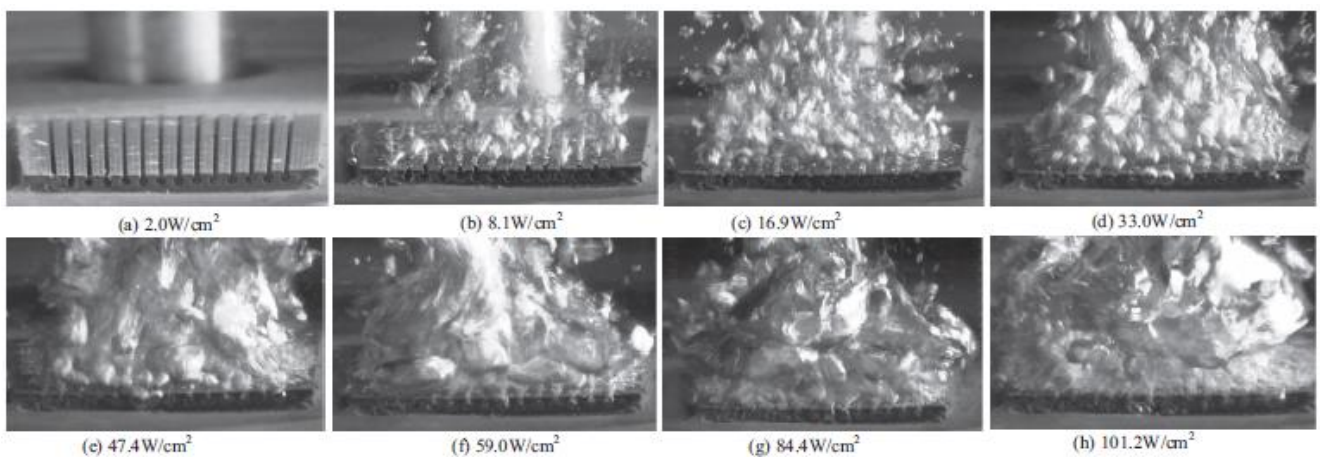

**Fig. 36.** Boiling visualizations of solid structures with reentrant channels ($\Delta T_{sub} = 3\ ^{0}C$, ethanol tests).

Mori et al. [17] have investigated the performance two layer structured honeycomb porous plate to increase the heat transfer performance in saturated pool boiling of water. In this study, a fine pores honeycomb porous plate (HPP) of very small thickness will be attached on the top of metal plate. A second HPP of larger pore diameter in comparison to first one will be stacked on the top of first HPP, attached on the metal plate. The first HPP is chosen to be as small as possible to reduce pressure drop, while the second HPP's pore diameter is made larger so that it can continuously supply liquid to prevent any dry out point or region on the plate. In their previous study, a single layer HPP was used to improve heat transfer coefficient and CHF. However, in this work, it is highlighted that the reduction of HPP thickness would reduced the pressure drop of both incoming liquid and outgoing vapour. At the same time, if the HPP thickness attached with the heated plate is made very small, then the heated plate will be dried when the vapour cover up the plate. To delay the occurrence of this incident, another HPP is attached on top of the bottom HPP. Here, it is also evident that gaps created between surface roughness and HPP structures have contributed significantly to escape vapour.

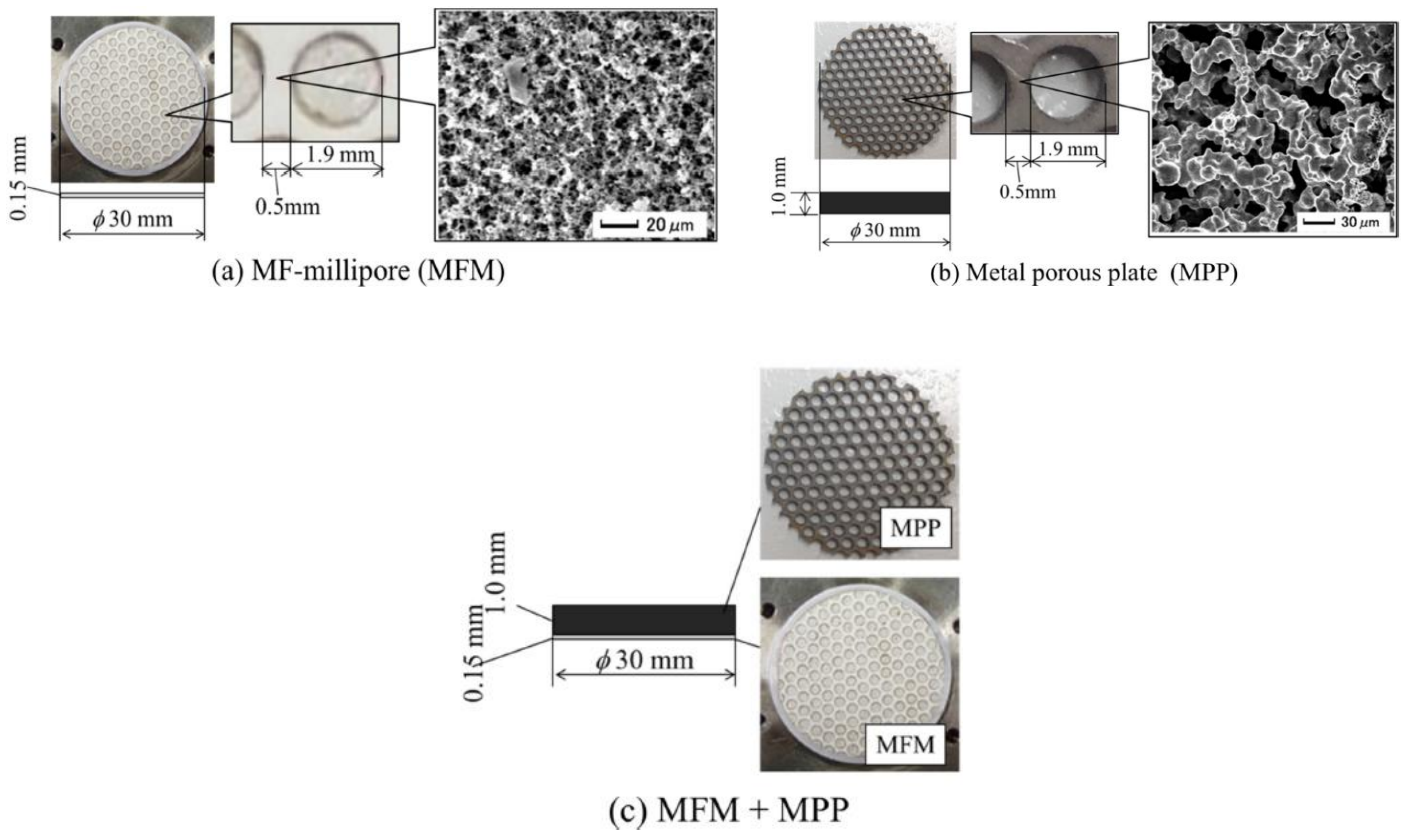

**Fig. 37.** Test porous plates.

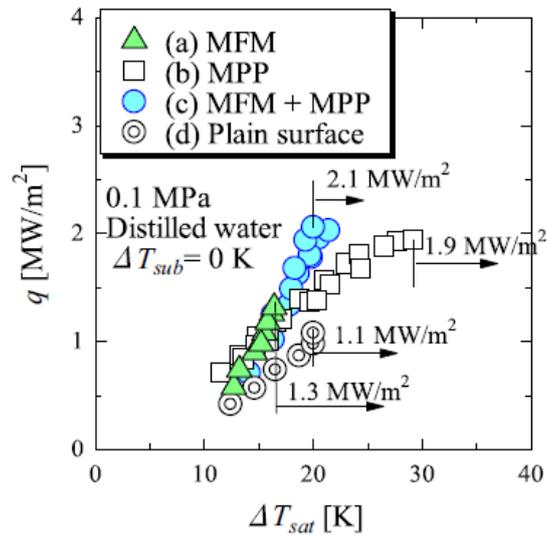

**Fig. 38.** Boiling curves for the (a) MFM, (b) MPP, (c) MFM + MPP, and (d) a plain surface.